\documentclass[prl,reprint,twocolumn,superscriptaddress,noshowpacs,notitlepage,longbibliography,10pt]{revtex4-1}%
\usepackage{graphicx,bm,times}
\usepackage{amsmath}
\usepackage{amsfonts}
\usepackage{amssymb}
\usepackage{color}
\usepackage[usenames,dvipsnames]{xcolor}
\usepackage[helvet]{sfmath}
\usepackage{hyperref}
\usepackage[normalem]{ulem}
\usepackage{soul}
\hypersetup{
	colorlinks = true,
	allcolors = {blue}
}
\begin{document}
\begin{titlepage}
\title{\Large{When superconductivity does not fear magnetism:\\Insight into electronic structure of RbEuFe$_4$As$_4$}}

\author{T.~K.~Kim}
\email[corresponding author:]{timur.kim@diamond.ac.uk}
\affiliation{Diamond Light Source, Harwell Campus, Didcot, OX11 0DE, United Kingdom}

\author{K.~S.~Pervakov}
\affiliation{Ginzburg Center for High Temperature Superconductivity and Quantum Materials Lebedev Physical Institute, Moscow 119991, Russia}

\author{D.~V.~Evtushinsky}
\affiliation{Laboratory for Quantum Magnetism, Institute of Physics, \'{E}cole Polytechnique F\'{e}d\'{e}rale de Lausanne, CH-1015 Lausanne, Switzerland}

\author{S.~W.~Jung}
\affiliation{Diamond Light Source, Harwell Campus, Didcot, OX11 0DE, United Kingdom}

\author{G.~ Poelchen}
\affiliation{Institut f\"{u}r Festk\"{o}rper- und Materialphysik, Technische Universit\"{a}t Dresden, D-01062, Dresden, Germany}
\affiliation{European Synchrotron Radiation Facility, 71 Avenue des Martyrs, Grenoble, France}

\author{K.~Kummer}
\affiliation{European Synchrotron Radiation Facility, 71 Avenue des Martyrs, Grenoble, France}

\author{V.~A.~Vlasenko}
\affiliation{Ginzburg Center for High Temperature Superconductivity and Quantum Materials Lebedev Physical Institute, Moscow 119991, Russia}

\author{V.~M.~Pudalov}
\affiliation{Ginzburg Center for High Temperature Superconductivity and Quantum Materials Lebedev Physical Institute, Moscow 119991, Russia}
\affiliation{National Research University Higher School of Economics, Moscow 101000, Russia}

\author{D.~Roditchev}
\affiliation{LPEM, ESPCI Paris, PSL Research University, CNRS, 75005 Paris, France}
\affiliation{Sorbonne Universite, CNRS, LPEM, 75005, Paris, France}
\affiliation{Moscow Institute of Physics and Technology, 141700 Dolgoprudny, Russia}

\author{V.~S.~Stolyarov}
\affiliation{Moscow Institute of Physics and Technology, 141700 Dolgoprudny, Russia}
\affiliation{N. L. Dukhov All-Russia Research Institute of Automatics, 127055 Moscow, Russia}

\author{D.~V.~Vyalikh}
\affiliation{Donostia International Physics Center (DIPC), 20018 Donostia-San Sebasti\'{a}n, Basque Country, Spain}
\affiliation{IKERBASQUE, Basque Foundation for Science, 48013 Bilbao, Spain}

\author{V.~Borisov}
\affiliation{Institut f\"{u}r Theoretische Physik, Goethe-Universit\"{a}t Frankfurt, Max-von-Laue-Strasse 1, D-60438 Frankfurt am Main, Germany}

\author{R.~Valent\'{i}}
\affiliation{Institut f\"{u}r Theoretische Physik, Goethe-Universit\"{a}t Frankfurt, Max-von-Laue-Strasse 1, D-60438 Frankfurt am Main, Germany}

\author{A.~Ernst}
\affiliation{Institut f\"{u}r Theoretische Physik, Johannes Kepler Universit\"{a}t, A 4040 Linz, Austria}
\affiliation{Max-Planck-Institut  f\"{u}r  Mikrostrukturphysik,  Weinberg  2,  D-06120  Halle,  Germany}

\author{S.~V.~Eremeev}
 \affiliation{Institute of Strength Physics and Materials Science, Russian Academy of Sciences, 634055 Tomsk, Russia}
 \affiliation{Tomsk State University, 634050 Tomsk, Russia}
 \affiliation{Saint Petersburg State University, 198504 Saint Petersburg, Russia}

\author{E.~V.~Chulkov}
\affiliation{Donostia International Physics Center (DIPC), 20018 Donostia-San Sebasti\'{a}n, Basque Country, Spain}
\affiliation{Saint Petersburg State University, 198504 Saint Petersburg, Russia}
\affiliation{Departamento de F\'{i}sica de Materiales UPV/EHU and Centro de F\'{i}sica de Materiales (CFM-MPC), Centro Mixto CSIC-UPV/EHU, 20080 Donostia-San Sebasti\'{a}n, Basque Country, Spain}

\maketitle
\end{titlepage}

\begin{abstract}
\normalsize{{
In the novel stoichiometric iron-based material RbEuFe$_4$As$_4$ superconductivity coexists with a peculiar long-range magnetic order of Eu\,4f states. Using angle-resolved photoemission spectroscopy, we reveal a complex three dimensional electronic structure and compare it with density functional theory calculations. Multiple superconducting gaps were measured on various sheets of the Fermi surface. High resolution resonant photoemission spectroscopy reveals magnetic order of the Eu\,4f states deep into the superconducting phase.
Both the absolute values and the anisotropy of the superconducting gaps are remarkably similar to the sibling compound without Eu, indicating that Eu magnetism does not affect the pairing of electrons.
A complete decoupling between Fe- and Eu-derived states was established from their evolution with temperature, thus unambiguously demonstrating that superconducting and a long range magnetic orders exist independently from each other.
The established electronic structure of RbEuFe$_4$As$_4$ opens opportunities for the future studies of the highly unorthodox electron pairing and phase competition in this family of iron-based superconductors with doping.}}
\end{abstract}
%\date{\today}

\maketitle

%\normalsize
%\setlength{\baselineskip}{20pt}
%%%%%%%%%%%%%%%%%%%%%%%%%%%%%%%%%%%%%%%%%%%%%%%%
\section{\normalsize{INTRODUCTION}}
Even after many years of intensive research, the mechanism of the electron pairing in iron based superconductors (IBSC) is still not well understood. It is known that emergence of superconductivity (SC) with doping is accompanied by a suppression of the structural and magnetic transitions observed in undoped parent compounds. Theoretical considerations suggest that the pairing mechanism in IBSCs might be due to spin-fluctuation exchange~\cite{Mazin_PRL_101_057003_(2008), Hirschfeld_Rep_Prog_Phys_74_124508_(2011), Chubukov_AnnRevCondMattPhys_3_57_(2012), Scalapino_RMP_84_1383_(2012)}.
In the case of weakly doped compounds the superconducting gap should have a $s^{\pm}$ symmetry. However, numerous studies of these materials  demonstrate that the physics of the pairing is more complex because of the multiorbital and multiband nature of low-energy fermionic excitations~\cite{Yi_npj_QantumMatt_(2017), Sprau_Science_357_75_(2017), Rhodes_PRB_98_180503(R)_(2018)}.
It turns out that both the symmetry and the structure of the order parameter result from a rather non-trivial interplay between spin-fluctuation exchange, intra-band Coulomb repulsion, and the momentum structure of the competing interactions~\cite{Chubukov_PRX_6_041045_(2016), {Kreisel_PRB_95_174504_(2017)}}.

In a newly discovered class of iron based superconductors, the so called ``1144” family~\cite{Iyo_JACS_138_3410_(2016)}, there are particular interesting examples of \emph{A}EuFe$_4$As$_4$  (\emph{A}=Rb,Cs) compounds where superconductivity coexists with unusual Eu-magnetic order~\cite{Kawashima_JPSJ_85_604710_(2016), Meier_npj_QantumMatt_(2017), Jackson_PRB_98_014518_(2018), Borisov_PRB_98_064104_(2018)}.
The structure of these compounds (see Fig.\ref{Fig_1}(a)) can be viewed as an intergrowth between undoped EuFe$_2$As$_2$ and heavily overdoped \emph{A}Fe$_2$As$_2$.
As a result, RbEuFe$_4$As$_4$ becomes intrinsically hole-doped, exhibiting superconductivity with a high transition temperature ${\sim}$36K.
Moreover, the evidence of in-plane ferromagnetic ordering of the Eu$^{2+}$ spins below ${\sim}$15K was given by magnetization measurements for both polycrystalline~\cite{Liu_PRB_93_214503_(2016)}  and single crystal samples~\cite{JinKe_Bao_CrystGrowthDes_18_3517_(2018), Smylie_PRB_98_104503_(2018)}.
The optical conductivity measurements~\cite{Stolyarov_PRB_98_140506(R)_(2018)} revealed a fully opened SC gap of about 5meV, while the inelastic neutron scattering (INS) measurements showed spin resonance at 18meV, and a three-dimensional helical antiferromagnetic order of the Eu atoms was determined from the neutron diffraction~\cite{Iida_PRB_100_014506(R)_(2019)}.
It has been even proposed that  this exotic helical magnetic structure is generated by superconductivity due to the interplay of Ruderman-Kittel-Kasuya-Yosida (RKKY) exchange interaction and macroscopic electromagnetic interaction between the superconducting and magnetic subsystems~\cite{Devizorova_PRB_100_104523_(2019), Koshelev_PRB_100_224503_(2019)}.
However, magnetic susceptibility and resistivity studies of RbEuFe$_4$As$_4$ under pressure found that the superconductivity onset is suppressed monotonically by pressure while the magnetic transition was enhanced at higher pressures~\cite{Jackson_PRB_98_014518_(2018)}.
Muon spin rotation ($\mu$SR) measurements of polycrystaline samples~\cite{Holenstein_arxiv_mSR} indicated that under hydrostatic pressure the superconducting transition temperature ${\rm T}_{\rm c}$  decreased, while the magnetic transition temperature  ${\rm T}_{\rm m}$ increased, suggesting no coupling between $s^{\pm}$ superconductivity typical for IBSC and exotic magnetic order of Eu.
A recent study of the magnetism and superconductivity in Ni-doped RbEuFe$_4$As$_4$ showed that upon doping the SC transition critical temperature ${\rm T}_{\rm c}$ decreased, but the Eu magnetic transition temperature ${\rm T}_{\rm m}$ is hardly affected~\cite{Willa_PRB_101_064508_(2020)}.
This suggests that Eu magnetism and superconductivity might be weakly dependent from each other.

In the present study we elucidate the interdependence of Fe-superconductivity and Eu-magnetism in this novel class of materials by combining experimental observations with theoretical calculations. In particular, we investigate the question whether the spatial proximity of the ferromagnetically ordered layer of Eu atoms can induce a non-trivial pairing of the Fe conduction electrons with exotic symmetry of the superconducting gap.

%%%%%%%%%%%%%%%%%%%%%%%%%%%%%%%%%%%%%%%%%%%%%%%%
\section{\normalsize{RESULTS}}
\subsection{\normalsize{Fermi surface topology}}
\begin{figure}[tb!]
\centering
\includegraphics[width=\linewidth]{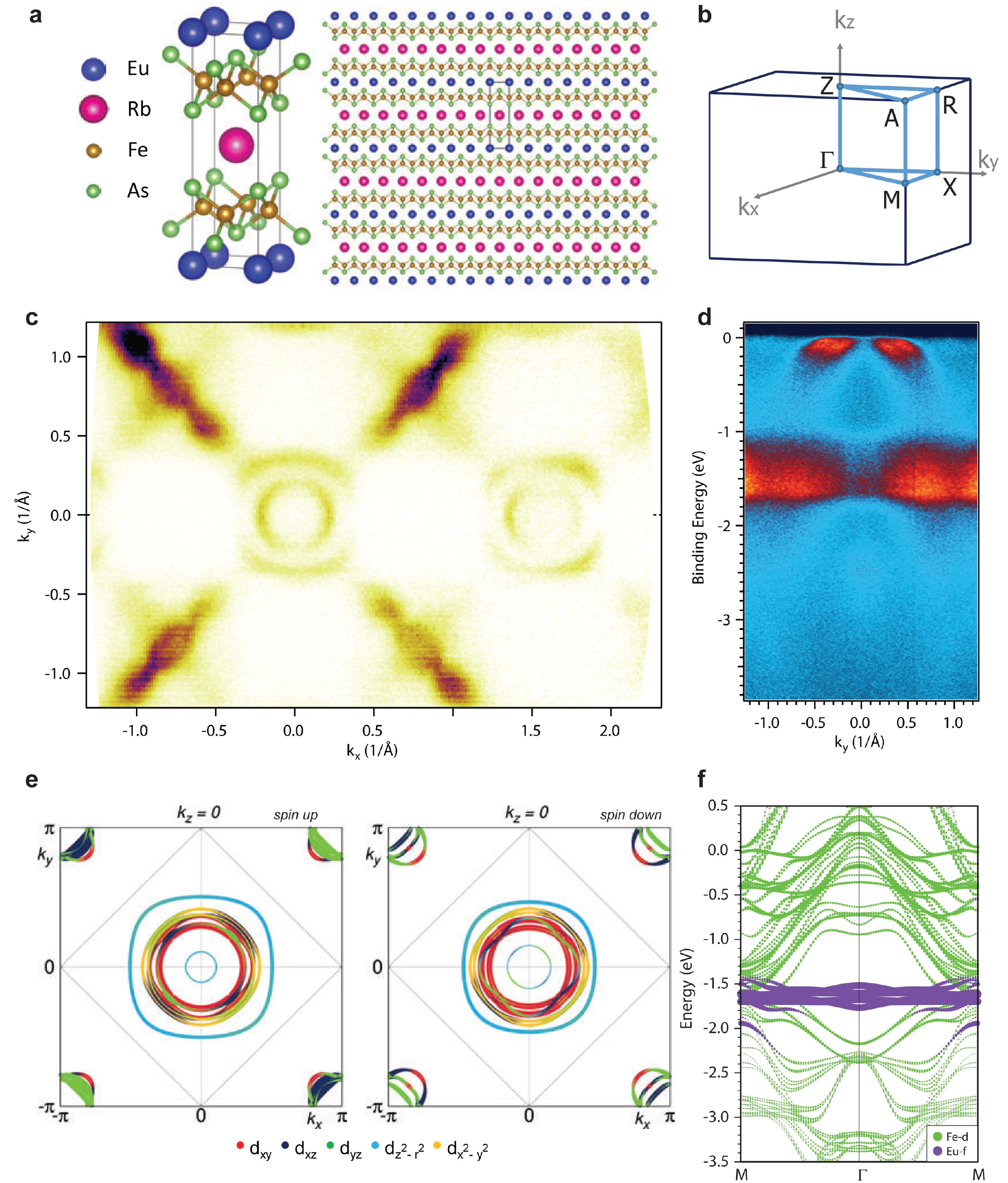}
\caption{\normalsize{\textbf{Structure and Fermi surface topology:}
(a) Layered crystal structure and (b) bulk Brillouin zone of RbEuFe$_4$As$_4$.
(c) $k_x$-$k_y$ Fermi surface map, measured using 70\,eV photons in linear vertical (LV) polarisation.
(d) Band dispersions in ${\rm M}$-${\Gamma}$-${\rm M}$ direction measured using 113\,eV photons for LV polarisation.  All data measured at 40K.
(e) Calculated Fermi surface for $k_z$=0 for FM configuration of Eu moments.
(f) Calculated band dispersions in ${\rm M}$-${\Gamma}$-${\rm M}$ direction showing Fe\,3d (green) and Eu\,4f (blue) bands.}}
\label{Fig_1}
\end{figure}

In Fig.\ref{Fig_1}(c), we present a Fermi surface (FS) map obtained at 70\,eV, corresponding to $k_z$=0 for the $\Gamma$ point where the hole pockets are largest.
The measured Fermi surface is similar to optimally hole-doped Ba$_{1-x}$K$_{x}$Fe$_2$As$_2$ (Ba``122") pnictides and consists of three hole-like pockets at the centre of the Brillouin zone (${\Gamma}$-${\rm Z}$ points in Fig.\ref{Fig_1}(b)) and propeller-like electron-like pockets at the corners of the the Brillouin zone (${\rm M}$-${\rm A}$ points in Fig.\ref{Fig_1}(b))~\cite{Zabolotnyy_Nature_2009}.
%}
High symmetry direction band dispersions measured in ${\rm M}$-${\Gamma}$-${\rm M}$ direction in Fig.\ref{Fig_1}(d) show the 4f electron emission from bulk Eu atoms, between 1.0 and 1.7\,eV below the Fermi level.
We note that there is no signature of 4f emission from surface Eu atoms, which would appear as an additional broad 4f signal at higher binding energies~\cite{Martensson_PRB_25_1446_(1982), Schneider_PRB_28_2017_(1983)}.
Apparently, the spectral pattern of Eu\,4f states lies far away from the Fe\,3d states that are crossing the Fermi level. This observation also suggests that it is hard to anticipate any interaction between Eu and Fe sublattices~\cite{Eu-Fe_distance}.

The experimentally observed results are in good agreement with DFT calculations for RbEuFe$_4$As$_4$ (see Fig.\ref{Fig_1}e,f). Like in many other known classes of iron-based superconductors, DFT overestimates the size of both hole and electron Fermi surfaces, leading to the so-called "red-blue" shift of the Fe\,3d bands: hole-like bands at the centre of the Brillouin zone moving to higher binding energies, while electron-like bands at the corners of the Brillouin zone moving to lower binding energies~\cite{Borisenko_NaturePhysics_2016, Zantout_PRL_123_256401_(2019), Bhattacharyya_PRB_102_035109_2020}.
This results in smaller size Fermi surfaces observed in experiment and in some cases to a corresponding increase of the density of states at the Fermi level due to the van-Hove singularity at the bottom of the electron pocket that moves toward zero energy at optimal doping~\cite{Kordyuk_vH}.

\subsection{\normalsize{Superconducting properties}}
\begin{figure}[tb!]
\centering
\includegraphics[width=0.85\linewidth]{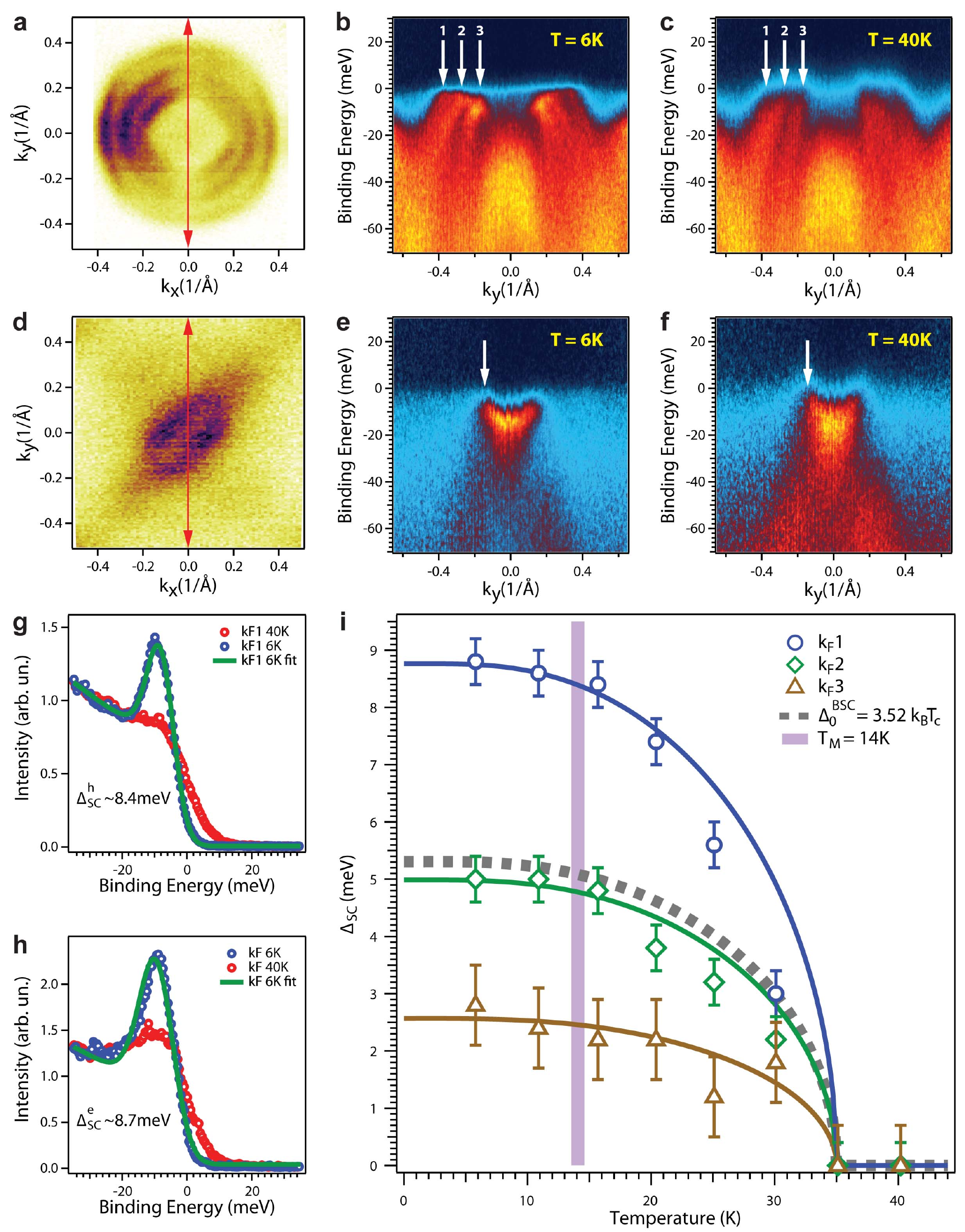}
\caption{\normalsize{\textbf{Superconducting properties:}
(a) Hole pocket Fermi surface at {\rm Z}-point (40K); % Z-point (27eV)
(b,c) Band dispersions at {\rm Z}-point below and above ${\rm T}_{\rm c}$ measured along high symmetry direction shown as red double arrow in panel (a), white arrows indicate corresponding ${k}_{\rm F}$ positions;
(d) Electron pocket Fermi surface at {\rm M}-point (40K); % M-point (38eV)
(e,f) Band dispersions at {\rm M}-point below and above ${\rm T}_{\rm c}$ measured along high symmetry direction shown as red double arrow in panel (d); white arrows indicate corresponding ${k}_{\rm F}$ positions;
(g,h) Photoemission spectra at equivalent ${k}_{\rm F}$ positions above and below ${\rm T}_{\rm c}$ together with the superconducting gap fit for hole and electron pockets;
(i) Temperature dependence of the superconducting gap obtained for three different Fermi surface sheets of the hole pocket at {\rm Z}-point.}}
\label{Fig_2}
\end{figure}

Using high resolution ARPES we have observed an opening of the superconducting gap on all sheets of the Fermi surface below ${\rm T}_{\rm c}$ (Fig.\ref{Fig_2}). The measured values of the gap at the hole pockets at the centre of the Brillouin zone are significantly varying from one sheet to another, with the largest gap opening on the most inner hole-like Fermi surface, similar to other Ba``122" pnictides~\cite{Hong_Ding_SCgap122, Evtushinsky_PRB_79_054517_(2009), Evtushinsky_PRB_89_064514_(2014)}.
We have also observed opening of the superconducting gap at the centre of the electron pocket. The obtained maximum value of the gap at the electron pocket is about $\sim$9\,meV according to the fit using the Dynes function~\cite{Dynes_PRL_41_1509_(1978)} and is similar to the largest gap value for the hole pocket (Fig.\ref{Fig_2}(g,h)). This is in a very good agreement with superconducting gap ARPES measurements for optimally doped Ba$_{1-x}$K$_{x}$Fe$_2$As$_2$~\cite{Hong_Ding_SCgap122, Evtushinsky_PRB_79_054517_(2009), Evtushinsky_PRB_89_064514_(2014)}.
These maximum gap values derived by ARPES are also in a very good agreement with published data from optical measurements of RbEuFe$_4$As$_4$~\cite{Stolyarov_PRB_98_140506(R)_(2018)}.

With increasing temperature, the superconducting gap gradually closes for all bands with classical BCS-type dependence, indicating a significant inter-orbital interaction of the Fe\,3d bands~\cite{Suhl_PRL_3_552_(1959), Moskalenko_SovPhysUsp_17_450_(1974), Daghero_SST_23_043001_(2010)}, similar to sibling compound CaKFe$_4$As$_4$~\cite{Mou_PRL_117_277001_(2016)} and Ba``122" family~\cite{Evtushinsky_PRB_89_064514_(2014), Kuzmicheva_JETP_Letters_107_42_(2018)}.
By measuring the temperature dependence of the superconducting gap at the hole pockets (Fig.\ref{Fig_2}(i)) within our energy resolution we do not observe any change in the magnitude of the gap at the Eu magnetic transition ${\rm T}_{\rm m}{\sim}$14K.
This result suggests that superconductivity in RbEuFe$_4$As$_4$ that involves Fe\,3d states near the Fermi level might be fully decoupled from the magnetism on the Eu sublattice.

%\subsection{Comparison of the electronic band structure for different magnetic ordering of Eu sub-lattice}
In order to investigate the effects of Eu\,4f magnetism on the Fe\,3d electronic states in the vicinity of the Fermi level we performed band structure calculations within DFT for several types of Eu magnetic moment ordering: ferromagnetic (FM), A-type antiferromagnetic ordering similar to EuFe$_2$As$_2$ (AFM180)~\cite{Herrero-Martin_PRB_80_134411_(2009), Xiao_PRB_80_174424_(2009)}, and for the recently proposed helical antiferromagnetic alignment of Eu moments (AFM90)~\cite{Iida_PRB_100_014506(R)_(2019)}.
\begin{figure}[tb!]
\centering
\includegraphics[width=\linewidth]{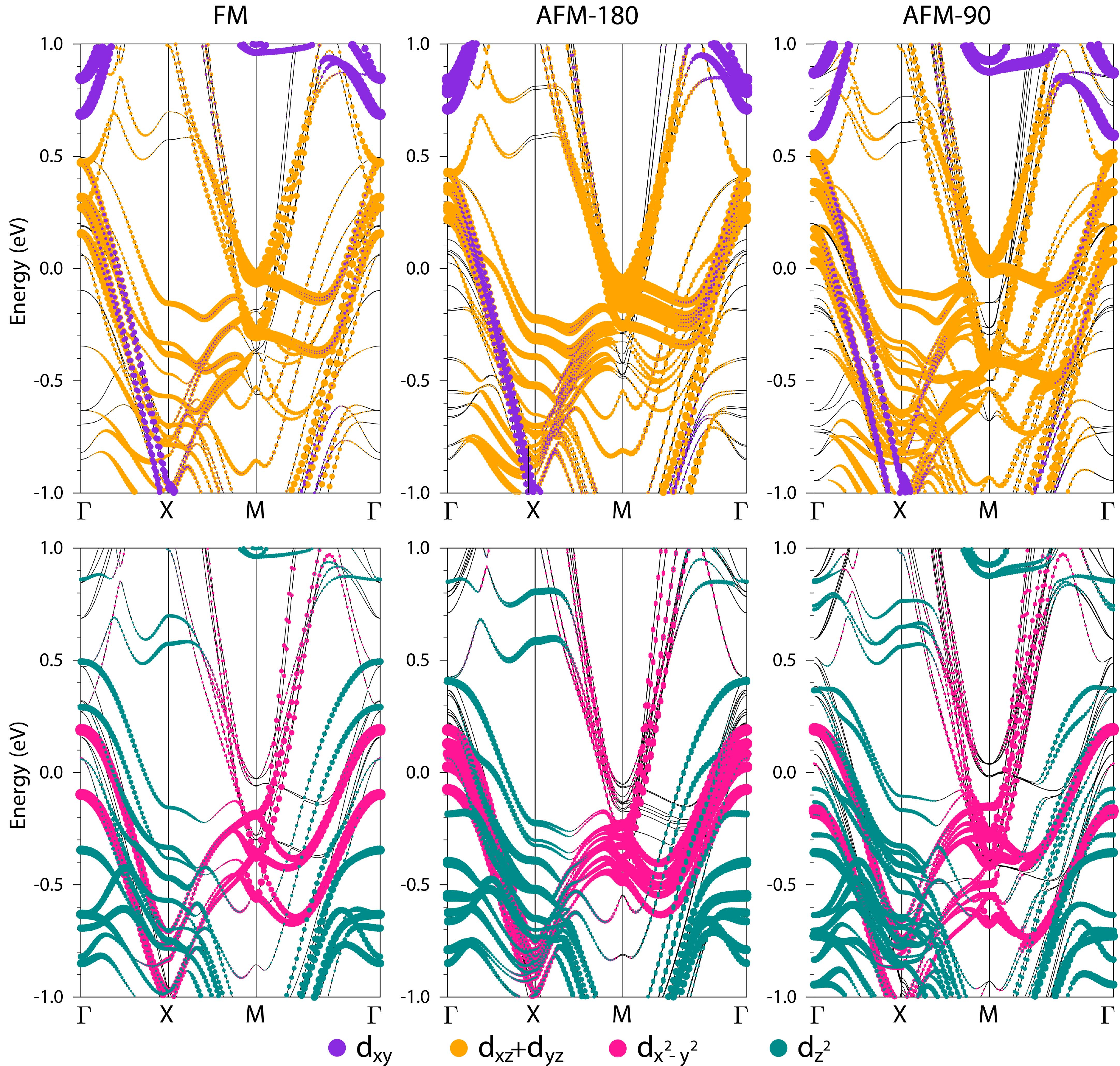}
\caption{\normalsize{\textbf{DFT electronic structure}
Band dispersions with the orbital character for ferromagnetic (left column), A-type antiferromagnetic (middle column) and helical antiferromagnetic (right column) Eu magnetic structures. Calculations done for 1, 2 and 4 formula units per cell correspondingly.}}
\label{Fig_3}
\end{figure}
As can be seen in Fig.\ref{Fig_3}, irrespective of magnetic ordering the bands forming electron ${\rm M}$-pockets at the Fermi level are almost completely determined by d$_{xz(yz)}$ Fe orbitals, whereas the bands forming hole $\Gamma$-pockets are contributed by a mixture of different  Fe\,3d orbitals.
In accordance with our calculations the basic band structure at the Fermi level and the orbital character of Fe\,3d bands are almost unaffected by different magnetic orderings of Eu\,4f magnetic moments.
Therefore the underlying spin-fluctuation-mediated interaction of Fe\,3d itinerant electrons responsible for superconducting pairing could be completely insensitive to the onset of the local magnetic order on Eu atoms.

\subsection{\normalsize{Magnetic properties}}
\begin{figure}[tb!]
\centering
\includegraphics[width=\linewidth]{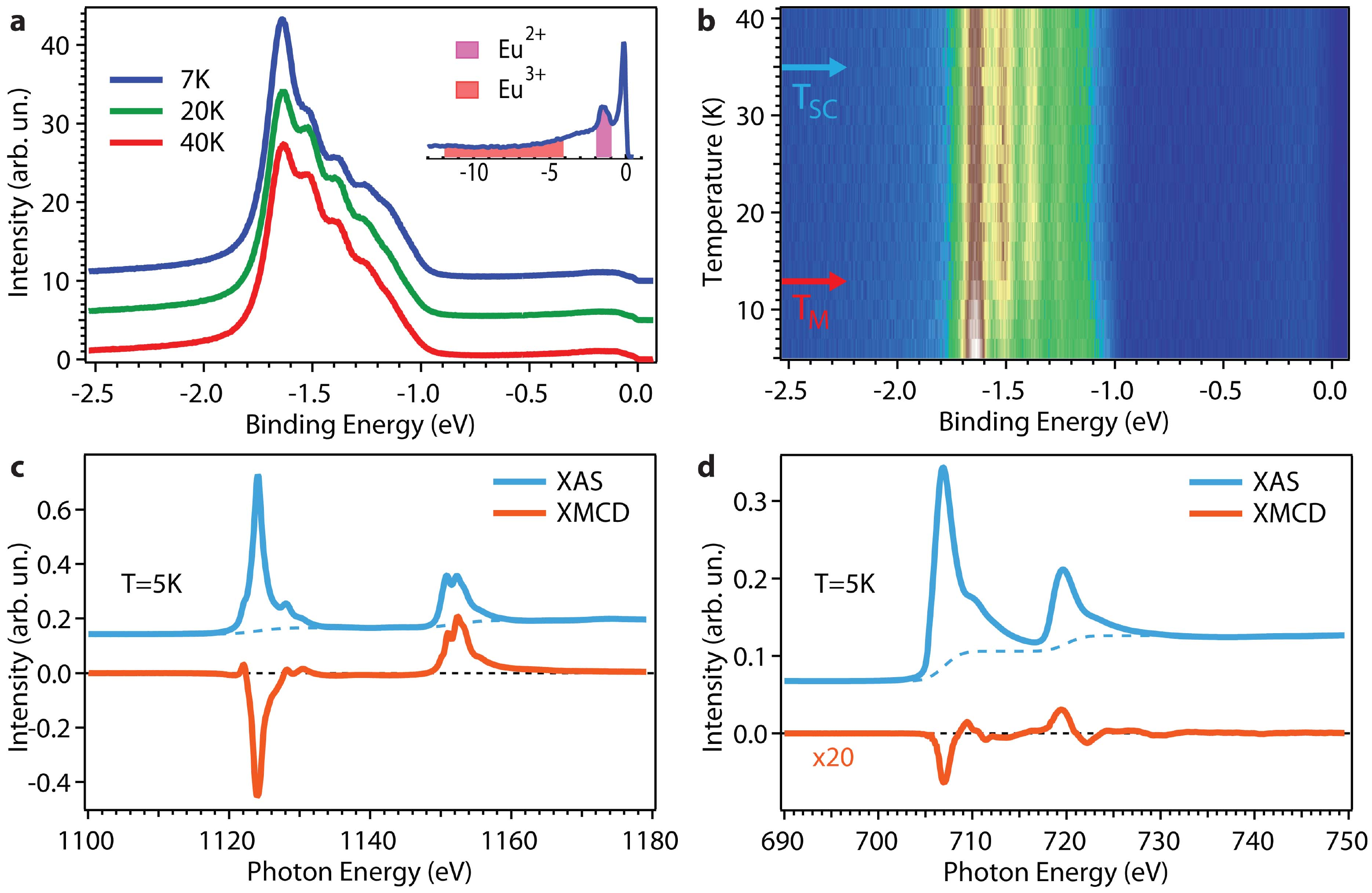}
\caption{\normalsize{\textbf{Magnetic properties:}
(a) The resonant Eu~4d$\rightarrow$4f photoemission spectra taken at 7, 30 and 40K with 142\,eV of photons and LV polarization; The 120\,eV survey scan shown as insert explicitly indicates no admixture of trivalent Eu state.
(b) 2D colour plot presentation of the temperature evolution of the divalent Eu\,4f resonant photoemission signal~\cite{Aurich}.
(c,d) XMCD spectra measured for Eu M$_{5, 4}$ and Fe L$_{3, 2}$ absorption edges.}}
\label{Fig_4}
\end{figure}

In order to explore the magnetic properties of the Eu sublattice, we performed resonant photoemission measurements at the Eu~4d$\rightarrow$4f threshold using 142\,eV  photons. This allows a resonant enhancement of the 4f emission from divalent Eu ions and gain insight exclusively into the spectral pattern of these states. The experiment was performed as a function of temperature going from the normal state (40K) via the superconducting state (20K) into the magnetic state (7K) as shown in Fig.\ref{Fig_4}(a,b).
Wide binding energy range valence band spectrum in insert of Fig.\ref{Fig_4}(a) clearly shows that Eu in RbEuFe$_4$As$_4$ is in pure Eu$^{2+}$ state with no contribution from Eu$^{3+}$ states.
Another essential point is that Eu\,4f sensitive ResPES spectra do not show any contribution from the surface 4f emission indicating that the analyzed spectral pattern originates from purely bulk Eu with no presence of the Eu$^{2+}$ at the surface.
The divalent Eu state with a 4f$^7$ electron configuration reveals a large and pure spin magnetic moment (${\rm J}$=${\rm S}$=7/2) of 7$\mu_{\rm B}$ which is responsible for the complex magnetic properties of Eu sublattice. Note that because the orbital moment of divalent Eu\,4f is equal to zero, this large magnetic moment will be rather insensitive to the crystal-electric field environment.
High resolution ResPES data for Eu\,4f states clearly show 4f$^7\rightarrow$4f$^6$ final-state Eu multiplet where the individual ${\rm J}$ components are well resolved (Fig.\ref{Fig_4}(a)) and in good agreement with both experiment and theoretical calculations~\cite{Gerken_JPhF_MPh_13_703_(1983),
  Starke_Mag_Dichr_(2000)}.
When following the evolution of the Eu\,4f spectral pattern with decreasing temperature (Fig.\ref{Fig_4}(b)), we can clearly see that there are no apparent changes in its shape and intensity when passing the onset of the superconducting order. However, there is a significant redistribution of the photoemission intensity between the individual ${\rm J}$-terms of the Eu\,4f multiplet below 14K. The latter can be explained as the appearance of the long-range ferromagnetic order of the 4f moments in the Eu layer~\cite{Starke_Appl_Phys_A_60_179_(1995), Starke_Mag_Dichr_(2000)}.
When the ferromagnetic order sets in due to exchange interaction, which couples neighboring 4f moments, the angular momenta assume long-range orientation leading to a preferred direction between incoming light and angular momentum. The latter implies a different excitation probability for the ferromagnetically ordered 4f states than for the paramagnetically ordered 4f states with nearly isotropic angular momentum orientation.

Furthermore, in Fig.~\ref{Fig_4}(c,d) we performed XMCD measurements to directly probe magnetic moments of both Eu and Fe atoms. Following standard sum rule analysis, we obtained a large magnetic moment 7.04$\pm$0.01$\mu_{\rm B}$ for Eu and a tiny magnetic moment 0.03$\pm$0.01$\mu_{\rm B}$ for Fe .
This value of the local magnetic moment on Eu is in close agreement with published transport data~\cite{Liu_PRB_93_214503_(2016), JinKe_Bao_CrystGrowthDes_18_3517_(2018), Smylie_PRB_98_104503_(2018)}.
The DFT calculated Eu magnetic moment of 6.97$\mu_{\rm B}$ is in excellent agreement with experiment, and is the same for any considered magnetic configuration.

To investigate magnetic properties of Fe sublattices, we used a disordered local moment (DLM) approach, in which magnetic moments of individual atoms are randomly orientated as in a paramagnetic state~\cite{Gyorffy1985, Staunton1985}. This gives the magnitude of Fe magnetic moments of 1.47$\mu_{\rm B}$. Note however that XMCD is sensitive only to the ordered component of the magnetic moment, which is very small, indicating a large fluctuating moment on the Fe sites.
FPLO calculations for the experimental structure with the FM order of Eu moments and ${\rm U}$=5~eV and ${\rm J}$=1~eV give the Fe moments of 1.7*10$^{-3}$ $\mu_{\rm B}$ and Eu moments of 7.13 $\mu_{\rm B}$.
Obtained value of the ordered Fe moment is consistent with XMCD measurements, while DLM shows the local moment which can be much larger.
Applying the magnetic force theorem~\cite{Liechtenstein1987} we found that the exchange interaction between the Eu und Fe moments is rather weak and negative indicating an antiferromagnetic coupling between the moments.
At the same time, our calculations have shown that AFM90 configuration for Eu moments is the most favorable. The total energy of the AFM180 is of 1.8\,meV/f.u. higher whereas the FM configuration is less favorable among considered being of 4.4\,meV/f.u. higher in energy respecting AFM90.

%%%%%%%%%%%%%%%%%%%%%%%%%%%%%%%%%%%%%%%%%%%%%%%%
\section{\normalsize{DISCUSSION}}
Using high resolution ARPES, we observed that the Fermi surface of RbEuFe$_4$As$_4$ consists of multiple hole-like and electron-like sheets, similar to other iron-based superconductors.
In this unique compound, the itinerant Fe\,3d electrons at the Fermi level are in proximity to the layers of ferromagnetically ordered large local magnetic moments of Eu atoms.
Therefore, the observed coexistence of Eu helical antiferromagnetic order with Fe superconductivity raises a widely discussed question about the possibility of non-trivial superconducting pairing in this material.
One of the most direct ways to answer the question about prevailing exotic pairing and its relation to co-existing Eu magnetic order is to probe the symmetry of the superconducting order parameter. RbEuFe$_4$As$_4$ is an intrinsically hole-doped iron-based superconductor with a high transition temperature ${\rm T}_{\rm   c}{\sim}$35K.
Our ARPES data show the opening of a full superconducting gap on all Fermi surface sheets below ${\rm   T}_{\rm c}$ with average gap value of $\sim$5\,meV, in good agreement with optical conductivity study~\cite{Stolyarov_PRB_98_140506(R)_(2018)}.
The fact that the temperature dependence of the various hole-like superconducting gaps follows a similar BCS-like relation for different sheets of the Fermi surface unambiguously points towards a significant interorbital coupling in this system~\cite{SCgap_interorbital}.
The Fermi surface topology as well as temperature dependence and anisotropy of the order parameter are very similar for RbEuFe$_4$As$_4$ and optimally doped Ba$_{1-x}$K$_{x}$Fe$_2$As$_2$. This suggests that superconductivity in both compounds is of the same origin and is consistent with theoretically proposed $s^{\pm}$ pairing due to the spin-fluctuations.

The temperature dependence of the hole-like superconducting gaps does not show any uncommon behaviour below the three-dimensional helical antiferromagnetic ordering of the Eu$^{2+}$ spins at 14K.
Moreover, our ARPES data show that weakly dispersing Eu\,4f states are at 1-1.7eV below the Fermi level, and our DFT calculations confirm that these localized states do not hybridize with itinerant Fe\,3d states and therefore do not contribute to the superconducting pairing directly.
Both Fe- and Eu-subsystems are almost decoupled magnetically, since the Fe\,3d bands and the Eu\,4f states are well separated in energy, despite structural proximity of Fe-As and Eu layers.
Besides, the calculations of the band structure in vicinity of the Fermi level for different types of magnetic orders of Eu moments establish that the topology and the orbital character of Fe\,3d bands are not constrained by the particular magnetic structure.
All these observations suggest that the iron-derived states near the Fermi level are completely independent from the localized Eu\,4f electronic subsystem.

Furthermore, we have also addressed the proposition~\cite{Devizorova_PRB_100_104523_(2019)} that superconductivity is responsible for the stabilisation of the helical antiferromagnetic ground state.
Using high resolution resonant photoemission measurements we not only observed the Eu\,4f final state multiplet structure, but, by measuring its temperature and polarisation dependence, confirmed in plane ordering of Eu$^{2+}$ localized moments below ${\rm T}_{\rm m}$.
These data also show that Eu~4f$^7$$\rightarrow$4f$^6$ final-state multiplet structure and therefore the magnetic arrangement does not change with the onset of superconductivity at ${\rm T}_{\rm c}$.
DFT calculations not only give correct binding energies of the Eu\,4f bands and correct magnetic moments of Eu, but also confirm that the helical magnetic structure is the most energetically favourable one.

The tiny magnetic moment of Fe obtained in our XMCD experiment is in agreement with results of M\"{o}ssbauer spectroscopy measurements~\cite{Albedah_2018}. These M\"{o}ssbauer   experiments also suggest that Fe-subsystem might not be magnetically ordered despite the long-range order of Eu-subsystem.
Using DLM approximation, we show that exchange interaction between Fe and Eu is small and negative.
Since existing DFT functionals can not describe correctly charge and spin fluctuations, we can not determine magnetic order in the Fe sublattice and assume here that the Fe moments are disordered.
The magnetism of iron in RbEuFe$_4$As$_4$ could be different from isostructural CaKFe$_4$As$_4$~\cite{Borisov_PRB_98_064104_(2018), Meier_npj_QantumMatt_(2017)} due to the possible biquadratic Eu-Fe coupling~\cite{Maiwald_PRX_8_011011_(2018)}.
Nevertheless, both magnetic RbEuFe$_4$As$_4$ and non-magnetic CaKFe$_4$As$_4$ compounds have similar Fermi surface topology~\cite{Mou_PRL_117_277001_(2016)}. In both cases the superconducting gaps for different Fermi surface sheets have no clear nodes and are roughly isotropic.
These observations once again demonstrate that superconductivity in RbEuFe$_4$As$_4$ is completely independent from the magnetic order on Eu .

RbEuFe$_4$As$_4$ is one of the few examples of the iron-based superconducting compounds with high transition temperature without additional doping.
The superconductivity in this material is enhanced for several reasons: due to the stoichiometric chemical composition there is no defect scattering detrimental for superconductivity; due to the charge carriers from Rb layers it has maximized density of states at the Fermi level beneficial for superconductivity; and it has no long-range magnetic order of iron moments, which is a well-known competitor of the superconductivity.
Moreover, a recent study of the doped RbEu(Fe$_{1-x}$Ni$_x$)$_4$As$_4$ system shows that by introducing extra electrons with Ni substitution at doping levels above $x\sim$0.07 the \textsl{ferro}magnetic superconductor with ${\rm T}_{\rm c}$ $>$ ${\rm T}_{\rm m}$ transforms into a superconducting \textsl{ferro}magnet with ${\rm T}_{\rm m}$ $>$ ${\rm T}_{\rm c}$~\cite{Liu_PRB_96_224510_(2017)}.
This doping induced interchange opens an unique opportunity to study the underlying change of electronic structure and possible existence of unconventional superconductivity.
Therefore, understanding the electronic structure of the parent compound RbEuFe$_4$As$_4$ becomes extremely important for any future studies of prospective superconducting magnets among iron-based pnictides.

To summarize, we have performed high resolution ARPES and ResPES studies of a newly discovered iron based superconductor RbEuFe$_4$As$_4$ with helical magnetic order. We observed three hole-like Fermi surface pockets around $\Gamma$-point and small electron-like pockets around ${\rm M}$-point, all formed by Fe derived bands. Our DFT calculations show that topology and the orbital character of the Fe\,3d band does not strongly depend on the particular magnetic ordering of the Eu\,4f states.
A full nodeless superconducting gap with BCS-like temperature dependence has been observed on both hole- and electron-like bands below ${\rm T}_{\rm c}$. In particular, no deviations in the temperature dependence of the order parameter have been detected below the magnetic transition ${\rm T}_{\rm m}$ despite of a clear indication for an in plane magnetic order of Eu$^{2+}$ localized moments. All these facts unambiguously indicate that Eu-magnetism and Fe-superconductivity are almost fully decoupled in RbEuFe$_4$As$_4$.

%%%%%%%%%%%%%%%%%%%%%%%%%%%%%%%%%%%%%%%%%%%%%%%%
%\setlength{\baselineskip}{20pt}
\section{\normalsize{METHODS}}
\subsection{Single crystal sample growth}
High quality single crystals of RbEuFe$_4$As$_4$ were grown by the self-flux technique. Superconducting and magnetic transition temperatures  ${\rm T}_{\rm c}{\sim}$35K and ${\rm T}_{\rm m}{\sim}$14K were confirmed by magnetisation measurements~\cite{Vlasenko_growth}.
\subsection{X-ray spectroscopy measurements}
High resolution angle-resolved photoemission spectroscopy (ARPES) and resonant photoemission spectroscopy (ResPES) measurements were performed at the I05 beamline at the Diamond Light Source, UK~\cite{Moritz_RevSciInstr_I05}. The photoelectron energy and angular distributions were analysed with a SCIENTA R4000 hemispherical analyser. The angular resolution was 0.2$^\circ$, and the overall energy resolution was better than 10\,meV for Fermi surface mapping and $\sim$2-3\,meV for SC gap measurements.
%In order to increase the contrast in Fig.\ref{Fig_4}(b) we have used a procedure described in V. Aurich, et al.\cite{Aurich1995}.
In order to increase the contrast in Fig.\ref{Fig_4}(b) we have used a procedure described in Ref.~\cite{Aurich1995}.
X-ray magnetic circular dichroism (XMCD) measurements were performed at the HECTOR end-station of the BOREAS beamline at the ALBA synchrotron radiation facility~\cite{Barla_JSR_(2016)_BOREAS_ALBA}.
Absorption spectra  in total electron yield mode have been recorded at the Fe L$_{3, 2}$ and Eu M$_{5, 4}$ edges in magnetic fields up to 6\,T and at 5K sample temperature, well below ${\rm T}_{\rm m}{\sim}$14K.
\subsection{First-principles calculations}
Density functional theory (DFT) calculations were performed considering various basis sets~\cite{Borisov_pssb_DFT_IBSC}: the projector augmented-wave (PAW) method for representation of   core electrons~\cite{Blochl_PAW1,Kresse_PAW2} as implemented in the VASP code~\cite{Kresse_VASP1,Kresse_VASP1}, the all-electron full-potential localized orbitals (FPLO) basis set code~\cite{Koepernik_FPLO}, and a full potential Green's function method within the multiple scattering theory~\cite{Geilhufe2015}.
Calculations were benchmarked with various codes. %Roser
The generalized gradient approximation (GGA-PBE)~\cite{Perdew1996} to the exchange-correlation potential was applied. To perform the calculations we have used the experimental crystal structure parameters~\cite{Liu_PRB_93_214503_(2016)}. The Eu\,4f states were treated employing the GGA+${\rm U}$ approach~\cite{Anisimov1991} within the Dudarev scheme~\cite{Dudarev1998}. The ${\rm U}_{\rm eff}$=${\rm U}$-${\rm J}$ value (where ${\rm U}$ and ${\rm J}$ are the effective on-site Coulomb and exchange interaction parameters, respectively) for the Eu\,4f states was chosen to be equal to 5.5\,eV. Using this ${\rm U}_{\rm eff}$ value we found a good agreement with the experimentally obtained binding energy of the Eu\,4f states of RbEuFe$_4$As$_4$~\cite{Vlad}.
%
%%%%%%%%%%%%%%%%%%%%%%%%%%%%%%%%%%%%%%%%%%%%%%%%
%\section{\normalsize{DATA AVAILABILITY}}
%The data analysed during the current study are available from the corresponding authors upon reasonable request.

%%%%%%%%%%%%%%%%%%%%%%%%%%%%%%%%%%%%%%%%%%%%%%%%
\section{\normalsize{ACKNOWLEDGEMENTS}}
%\begin{acknowledgments}
We thank Dr. Matthew Watson for his critical reading of the manuscript.
We thank Diamond Light Source for access to beamline I05 (proposal numbers SI15074, SI19041) that contributed to the results presented here.  Work was done using equipment from the LPI Shared Facility Center and support by the Russian Scientific Foundation (RSF project no. 20-12-00349).  The support by Tomsk State University competitiveness improvement program (No. 8.1.01.2018), the Saint Petersburg State University (Grant ID 51126254), and the Fundamental Research Program of the State Academies of Sciences (line of research III.23.2.9) is gratefully acknowledged.
RV acknowledges funding by the Deutsche Forschungsgemeinschaft (DFG) TRR 288 (project A05). VB thanks the Goethe University Frankfurt for computational resources and Daniel Guterding for providing the Fermi surface plotting software.
KK thanks M. Valvidares, J. Herrero, H. B. Vasili, S. Agrestini and N. Brookes for their support during the XMCD experiment.
DVV also acknowledge support from the Spanish Ministry of Economy (MAT-2017-88374-P).
%\end{acknowledgments}
%$\hspace{1pt}$

%%%%%%%%%%%%%%%%%%%%%%%%%%%%%%%%%%%%%%%%%%%%%%%%
%\section{\normalsize{AUTHOR CONTRIBUTIONS}}
%Single crystal samples were grown and characterized by K.S.P., V.A.V. and V.M.P
%ARPES and ResPES experiments were performed by T.K.K., D.V.E., S.W.J., G.P. and D.V.V.
%XMCD experiments were performed by K.K.
%DFT calculations were done by V.B., S.V.E. and A.E. led by R.V. and E.V.C.
%The obtained results were intensively discussed with V.S.S. and D.R.
%T.K.K. wrote the manuscript with input from all co-authors.
%
%%%%%%%%%%%%%%%%%%%%%%%%%%%%%%%%%%%%%%%%%%%%%%%%
\newpage
\section{\normalsize{REFERENCES}}
\bibliographystyle{naturemag}
%\bibliography{RbEuFe4As4}

\begin{thebibliography}{99}
\bibitem{Vlad}
\bibinfo{note}{{The binding energy of the Eu\,4f states in FPLO changes from
  -1.3 to -2.5\,eV when the ${\rm U}$ parameter is varied between 5 and 7\,eV,
  while keeping ${\rm J}$ = 1\,eV}}.
\end{thebibliography}

\end{document}